\documentclass[
	twocolumn,
	showpacs,
	preprintnumbers,
	amsmath,
	amssymb,
	aps,
	prl]{revtex4-1}

\usepackage{graphicx}
\usepackage{upgreek}
\usepackage{color}
\usepackage{xspace}
\usepackage{textcomp}

\newcommand{\perovskite}[2]{#1#2O$_3$\xspace}
\newcommand{\BBO}{\perovskite{Ba}{Bi}}
\newcommand{\BPO}{\perovskite{Ba}{Pb}}
\newcommand{\STO}{\perovskite{Sr}{Ti}}
\newcommand{\LAO}{\perovskite{La}{Al}}
\newcommand{\BBPO}{BaBi$_{1-x}$Pb$_x$O$_3$\xspace}
\newcommand{\BBPOF}{BaBi$_{0.25}$Pb$_{0.75}$O$_3$\xspace}
\newcommand{\BKBO}{Ba$_{1-x}$K$_x$BiO$_3$\xspace}
\newcommand{\eg}{e.\,g.\xspace}
\newcommand{\TBKT}{\ensuremath{T_\mathrm{BKT}}\xspace}
\newcommand{\Tc}{\ensuremath{T_\mathrm c}\xspace}
\newcommand{\Tcip}{\ensuremath{T_\mathrm{c,ip}}\xspace}
\newcommand{\scm}{cm$^2$\xspace}
\newcommand{\Hcpd}{\ensuremath{H_\mathrm{c2,\perp}}\xspace}
\newcommand{\Hcc}{\ensuremath{H_\mathrm{c2}}\xspace}
\newcommand{\Hcpl}{\ensuremath{H_\mathrm{c2,\parallel}}\xspace}
\newcommand{\dsc}{\ensuremath{d_\mathrm{sc}}\xspace}

\begin{document}

\title{Interface-Driven Two-Dimensional Superconductivity\\
 in Bilayers of \BBO and \BPO}

\author{B.~Meir}
\author{S.~Gorol}
\author{T.~Kopp}
\author{G.~Hammerl}

\affiliation{
Experimental Physics VI\\ 
Center for Electronic Correlations and Magnetism, Institute of Physics, 
University of Augsburg, 86135 Augsburg, Germany}

\date{\today}

\begin{abstract}
Besides the cuprate superconductors lead-doped barium bismuthate still attracts 
intense interest concerning its relatively high critical temperature and the 
nature of its superconductivity in vicinity to a charge density wave ordered 
topological insulator. 
Here we present bilayers of barium bismuthate and barium leadate and find 
two-dimensional superconductivity at their interfaces well described by 
current-voltage characteristics compatible with a 
Berezinski\u{i}-Kosterlitz-Thouless transition.
Magneto-transport studies independently prove the two-dimensional character of 
the superconducting properties in these bilayers. 
The particular dependence of the superconducting transition temperature 
on the thickness of the barium leadate top layer suggests the formation of 
the superconducting state to be strain-induced through the interface.
\end{abstract}

\maketitle

Interfaces in oxide heterostructures serve as an inexhaustible source for the 
creation of new electronic phases well constrained on the nanoscale 
\cite{mannhart:2010}. Specifically, superconductivity confined to an interface
is of particular importance not only because it is easy to tune by external 
fields but especially because the interface electron system results from an 
electronic reconstruction in dependence on bulk states with their versatile 
ground states~\cite{hesper:2000,okamoto:2004,comment:1,zubko:2011}.  
The interface electron structure is controlled by bulk materials properties but 
also by the symmetry breaking at the edge of the respective bulk materials. The 
prominent oxide heterostructure of the two band insulators \STO (STO) and \LAO 
(LAO) hosts such an interface-confined electron 
liquid~\cite{ohtomo:2004,ohtomo:2006,breitschaft:2010}
showing two-dimensional (2D) superconductivity at temperatures below 
$\approx$\,300\,mK~\cite{reyren:2007,caviglia:2008}. Structural inversion 
symmetry breaking at the interface introduces a sizable Rashba spin-orbit 
coupling~\cite{caviglia:2010}.
Also cuprate bilayers present striking features of their superconducting 
states: a high-\Tc state confined to 2--3\,nm is formed between insulating 
La$_2$CuO$_4$ and metallic La$_{1.55}$Sr$_{0.45}$CuO$_4$~\cite{gozar:2008}, 
both of them not superconducting. Moreover, heterostructures built from the 
insulating oxides CaCuO$_2$ and STO allow for a superconducting 
interface~\cite{dicastro:2015}. 

Up to now, only few superconducting oxide interfaces have been realized. However
it is an appealing challenge to identify further superconducting bilayers  
because different perovskite parent compounds account for novel characteristics 
of electronic states in proximity to the interface.
For example, \BBO (BBO) is predicted to be a large-energy gap topological 
insulator due to spin-orbit coupling (when electron doped)~\cite{yan:2013}, 
whereas the related compound \BPO (BPO) is a bulk metal \cite{ikushima:1966}. 
In fact, both BBO and BPO are expected to preserve a ``hidden'' topological
insulator phase when electron or hole doped~\cite{li:2015,comment:2}. 
Moreover, lead-doped bulk BBO is well-known to be 
superconducting~\cite{sleight:1974}. That finding now incites the question of 
whether the interface between BBO and BPO is a 2D superconductor. 
If this is the case, do the topological surface states generate Majorana 
fermions~\cite{fu:2008} and is that modulated with the 
Berezinski\u{i}-Kosterlitz-Thouless (BKT) transition which by itself is 
topologically driven? 
The potential realization of topologically protected edge states is certainly a 
strong motivation to explore the interfacial electronic system of BBO/BPO in 
detail, although here we keep our focus exclusively on establishing the 
formation of a 2D superconducting phase in bilayers of BBO and BPO.

BBO crystallizes at room and low temperatures in perovskite related 
monoclinic structures \cite{sleight:2015}. The underlying octahedral distortion 
leads to symmetric breathing-mode displacements forcing the Bi$^{3+}$ and 
Bi$^{5+}$ ions to distinct crystal sites forming a bond-order charge density 
wave (CDW)~\cite{cox:1976,cox:1979,varma:1988,kim:1996}.
The charge-ordered state can be altered by, \eg, substituting the metal cations 
in BBO giving rise to tremendeous changes in the structural and electronic 
properties of this very special perovskite: in homogeneosly doped bulk samples 
of \BBPO (BBPO) three-dimensional superconductivity is observed for doping 
ranges of $0.65<x<0.95$ with a maximum superconducting transition temperature 
of $\Tc\approx13$\,K for $x$ near 0.75 \cite{sleight:1975,thanh:1980, 
batlogg:1984,climent:2011,giraldo-gallo:2012}. 
The onset of superconductivity is attributed to the formation of a bimorphic 
phase consisting of orthorhombic and tetragonal polymorphs in 
BBPO~\cite{climent:2011,giraldo-gallo:2012,balachandran:2013,giraldo-gallo:2015}. 
By substituting barium by potassium, bulk samples of \BKBO yield maximum 
critical temperatures of $\Tc\approx30$\,K for a potassium concentration of 
$x=0.4$~\cite{cava:1988}. The high-\Tc superconductivity of these compounds was 
recently linked to correlation-enhanced electron-phonon 
coupling~\cite{yin:2013}.
After the discovery of superconductivity in bulk samples, thin films of BBPO 
were soon realized by different growth techniques including
sputtering~\cite{gilbert:1978,suzuki:1980,hidaka:1983} and pulsed laser 
deposition (PLD) \cite{zaitsev:1983}. 

Inspired by the discovery of a superconducting state at LAO/STO interfaces, 
we investigated BBO/BPO bilayers (BLs) in expectation to identify 2D 
superconductivity non-existent in both parent compounds. 
Thin films of nominal \BBPOF, BBO, BPO, and BBO/BPO BLs were grown by PLD using 
commercially available, stoichiometric targets with purities of at least 99.9\% 
at maximum reachable density.
We used single crystalline (001) oriented STO crystals as 
substrates, which we cleansed in acetone and isopropyl prior to deposition. 
HF buffering STO substrates \cite{kawasaki:1994,koster:1998} had no significant 
impact on the results. 
The deposition temperatures were $\approx$\,635\textcelsius{} for \BBPOF thin 
films and $\approx$\,552\textcelsius{} for BBO, BPO thin films and BBO/BPO BLs 
controlled by laser-heating and monitored by pyrometers. The background 
pressure of pure oxygen was kept mass-flow controlled at $\approx$\,1\,mbar 
during growth.
Our PLD system is equipped with a KrF laser having a fluency of $\approx$\,2\,J/\scm.
We used laser pulse energies in the range of 550--750\,mJ and laser pulse 
frequencies of 3--5\,Hz. After deposition the samples were cooled to 
$\approx$\,400\textcelsius{} within three minutes and annealed at a background 
pressure of oxygen of $\approx$\,400\,mbar for at least 17~minutes before the 
vacuum chamber was evacuated again. 

For \BBPOF thin films we used 1000 laser pulses (LP) resulting in film 
thicknesses of $\approx$\,420\,nm. For thin films of BBO and BPO we
used 50--150~LP corresponding to film thicknesses of $\approx$\,25--75\,nm for 
BBO and  $\approx$\,17--50\,nm for BPO.
Concerning the BBO/BPO BLs we kept the thickness of the BBO starting layer 
constant (always 100~LP) and only varied the number $N$ of LP of 
the BPO top layer.
 
\begin{figure}
\centering
\includegraphics[width=\columnwidth]{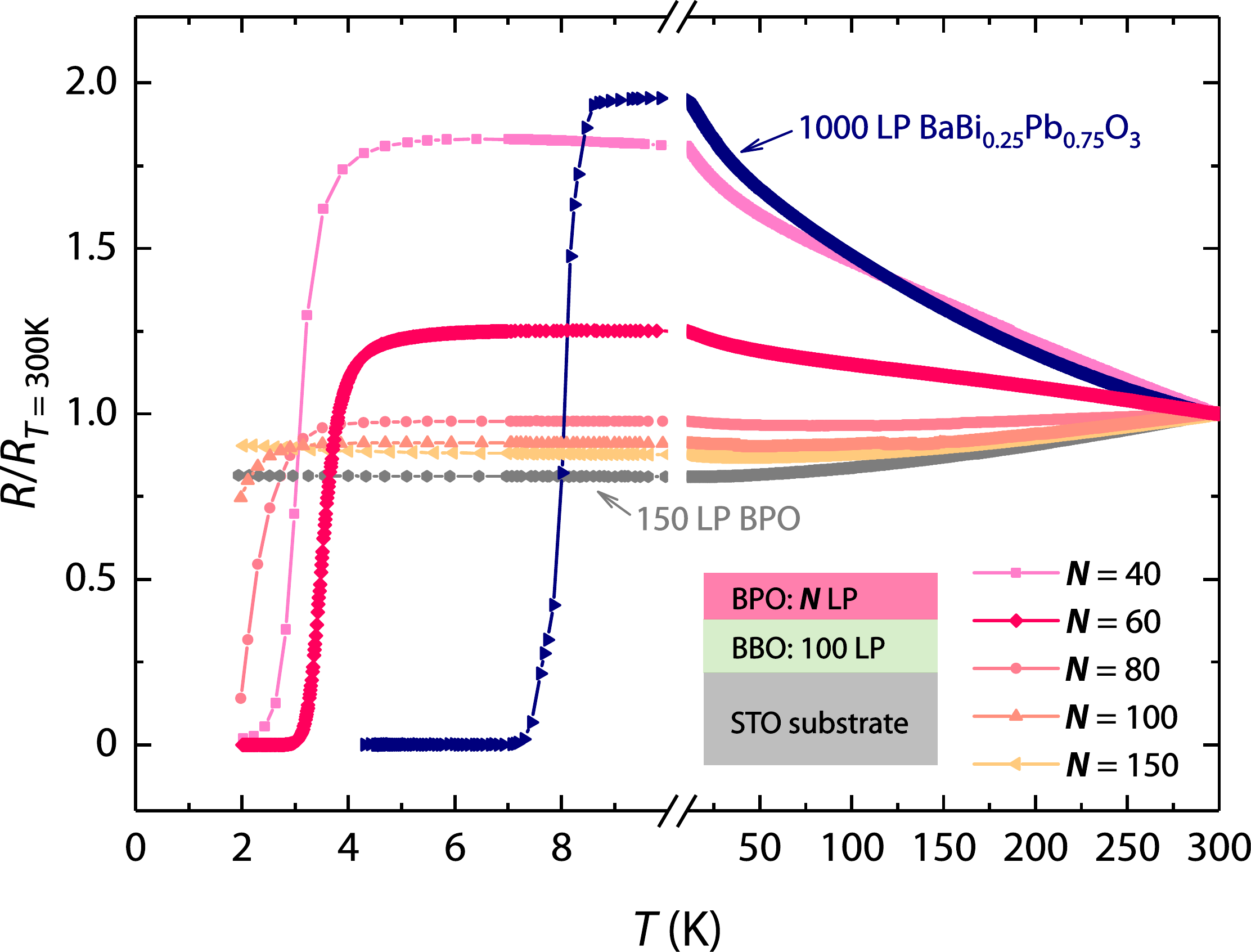}
\caption{Normalized resistances of pulsed laser deposition (PLD) grown thin 
films of \BPO (BPO) and \BBPOF and  \BBO (BBO)/BPO bilayers (BLs) as function 
of temperature. Whereas BBO thin films are highly insulating (not shown), BPO 
thin films show metallic behaviour (gray hexagons). Thin films of nominal 
\BBPOF are superconducting at low temperatures. BLs with a fixed thickness of 
BBO (100 laser pulses (LP)) and variable thickness of BPO top layer (controlled 
by the number of LP $N$) undergo superconducting transitions (samples with 
$N=\{40, 60, 80, 100\}$).}
\label{RT}
\end{figure}

All grown \BBPOF thin films show an increase in resistance with decreasing 
temperature and a superconducting transition at $\Tcip\approx8$\,K, which is 
determined at the inflection point (ip) of $R(T)$ at the transition (see 
Fig.~\ref{RT}, blue filled triangles facing right).

BBO and BPO thin films display peaks consistent with only (00$l$) oriented 
planes in routinely taken x-ray diffraction (XRD) experiments. BBO thin films 
are insulating with resistances ranging from M$\Omega$ to G$\Omega$ over the 
whole accessible temperature range. Pure BPO thin films instead turn out to be 
metallic (see~Fig.~\ref{RT}, gray hexagons).

BBO/BPO BLs reveal only peaks classified by (00$l$) orientation in XRD studies 
separated in individual reflexes for BBO and BPO, respectively. Peaks assigned 
to a homogeneously doped \BBPO phase could not be identified (see Supplemental 
Material~\cite{supplement}, Fig.~S1). BLs with $N=30$ are semiconducting at all 
accessible temperatures, BLs with $N=40$ show semiconductor-to-superconductor 
transitions with $\Tcip\approx3.0$\,K (see Fig.~\ref{RT}, pink filled squares). 
An increase of the BPO thickness ($N=60$) shifts the superconducting transition
temperature to values as high as $\Tcip=3.5$\,K (see Fig.~\ref{RT}, red 
diamonds). Interestingly, with further increase of $N$ we observe a dramatic 
reduction of the critical temperature ($N=\{80,100\}$) and no superconducting 
transition for samples with $N=150$. The observation of an optimal BPO 
thickness to reach maximum \Tc and the suppression of superconductivity beyond 
this thickness indicates that the BL superconductivity is related to the 
interface (see below) and may also be dependent on the thickness of the BBO 
starting layer (in the case of atomically thin BBO layers which is not 
investigated here).

\begin{figure}
\centering
\includegraphics[width=\columnwidth]{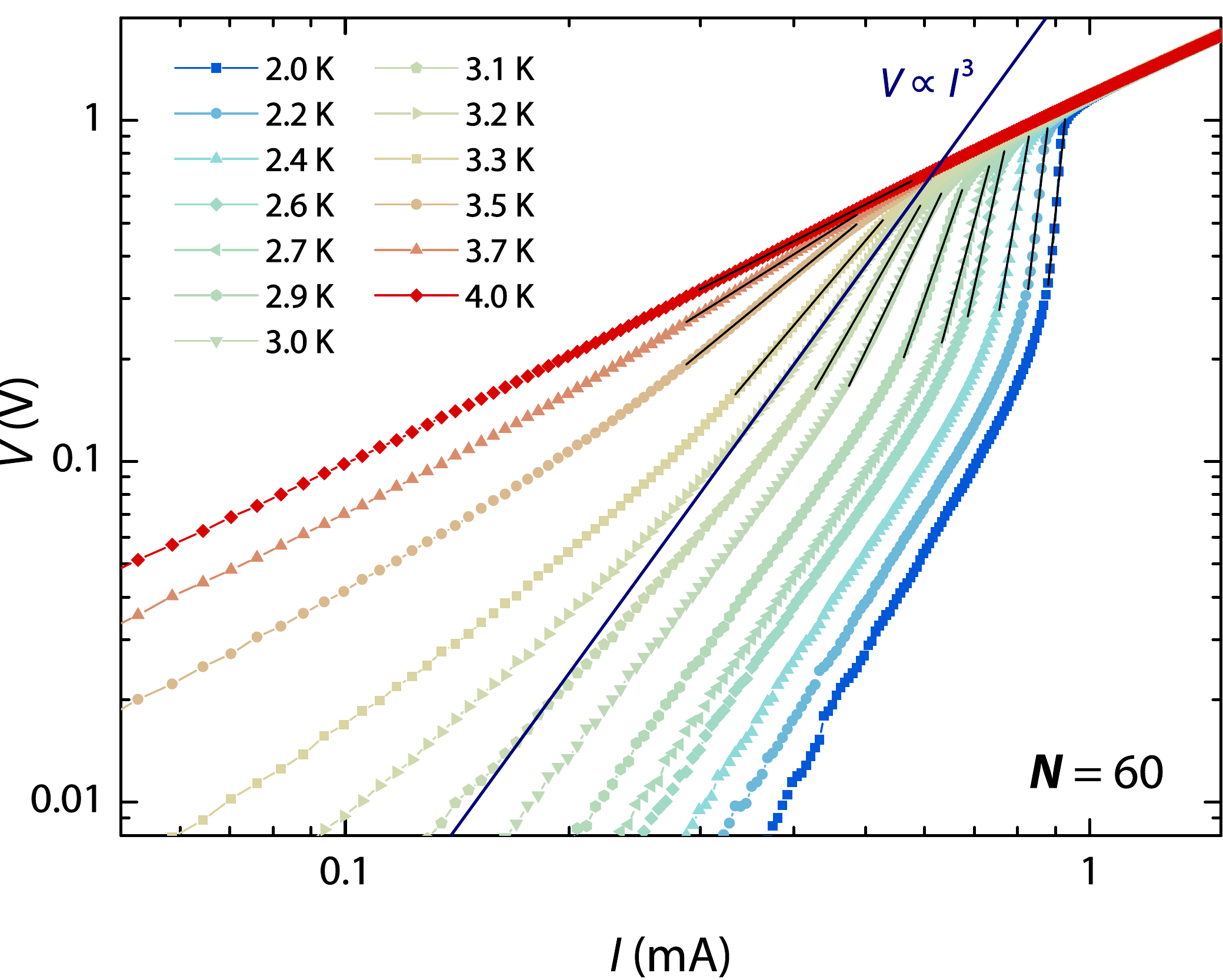}
\caption{Measured current-voltage (IV) characteristics of a BBO/BPO BL 
($N=60$) at temperatures ranging from 2\,K (blue) to 4\,K (red). Straight black 
lines are polynomial fits following the measured data directly at the 
transition. At 3.2\,K the IV curve is best represented by a cubic fit (straight 
blue line) identifying 2D superconductivity following a 
Berezinski\u{i}-Kosterlitz-Thouless (BKT) transition. The deviation of the fit 
for small currents is attributed to finite size effects~\cite{medvedyeva:2000}.}
\label{BKT}
\end{figure}

If the superconducting transition is really driven by the interface, then a 
likely scenario is that a 2D superconducting sheet is formed within the
bilayered samples. Such behavior is described by a BKT 
transition~\cite{berezinskii:1971,berezinskii:1972,kosterlitz:1973,kosterlitz:1974}. 
A convenient experimental method to characterize the transition into the 
algebraically ordered state is the familiar progression of current-voltage (IV) 
characteristics taken at temperatures around \Tc: the applied voltage $V$ 
breaks vortex-antivortex pairs and yields at $T=\TBKT$ the well-known 
$V=I^{\alpha}$ behavior with $\alpha=3$, whereas the current $I$ responds 
linear in $V$ for a state with free vortices above 
$\TBKT$~\cite{halperin:1979}. 
A direct validation of this observation can be made by analyzing the $R(T)$ 
dependence. For a BKT transition, the resistance 
follows~$R\propto R_\mathrm{n}\exp(-b/\sqrt{t})$~\cite{kosterlitz:1973} 
near $\TBKT$ with $b$ being a material parameter in the order of 
unity~\cite{halperin:1979} and $t=T/\TBKT-1$. 
Our data yield $b=1.22$ and $\TBKT=3.26$\,K matching the value 
$\TBKT=3.19$\,K obtained independently from the IV measurements (see 
Supplemental Material~\cite{supplement}, Fig.~S2). The perfect agreement of our 
results (see Fig.~\ref{BKT}) with the BKT predictions signifies 
unambiguously that indeed a 2D superconducting sheet is formed through the 
interface. The ohmic regime observed below $\TBKT$ for small currents is 
attributed to be responsible for the deviations from the fits due to finite 
size effects~\cite{medvedyeva:2000}.

\begin{figure}
\centering
\includegraphics[width=\columnwidth]{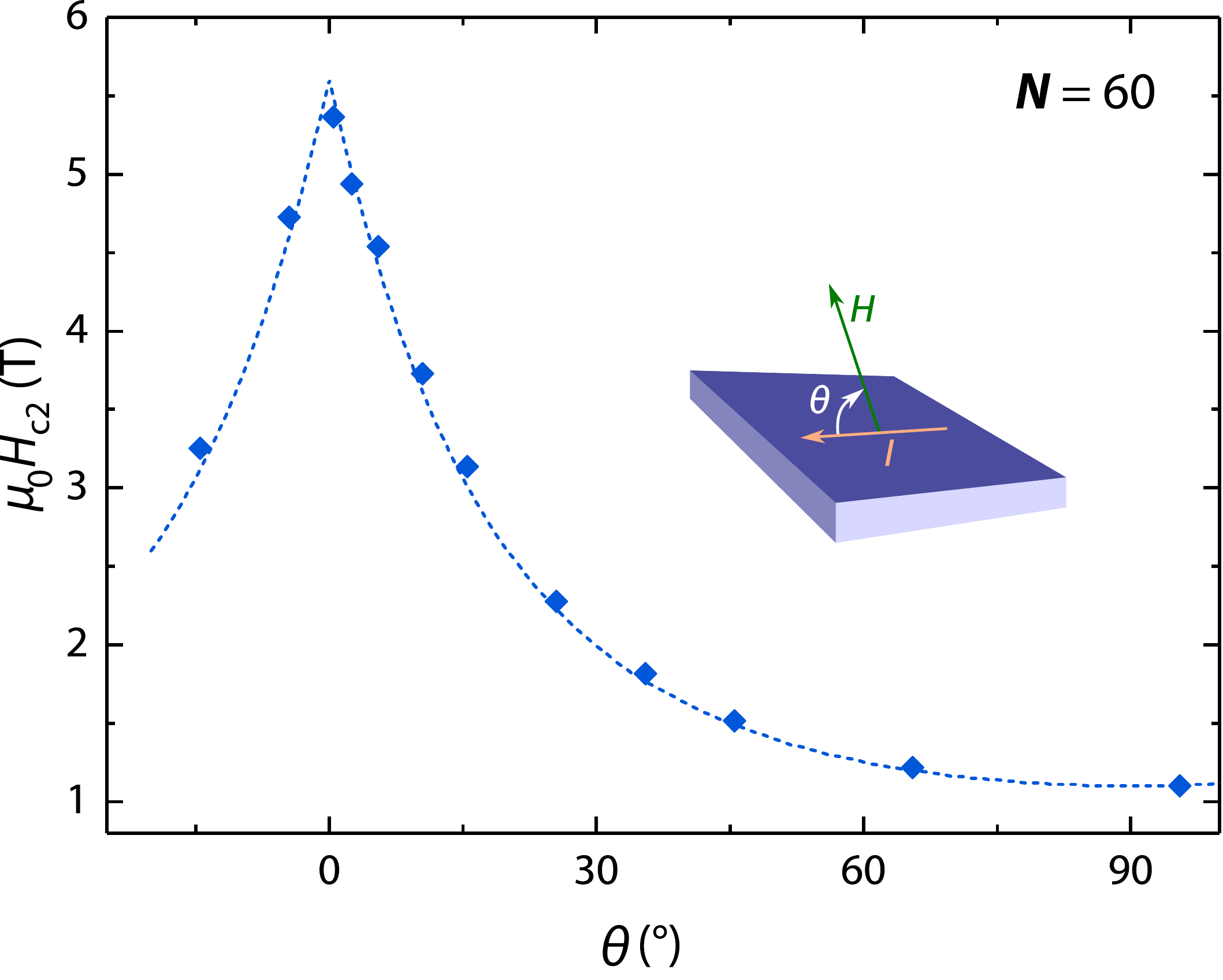}
\caption{Observed dependence of the upper critical magnetic field 
$H_\mathrm{c2}$ of a superconducting BL ($N=60$) on the direction $\theta$ of 
the externally applied swept magnetic field (filled diamonds). The data are 
well represented by Eq.~(\ref{tinkhamrelation}) (dashed line) which applies to 
a 2D superconducting sheet in the GL regime~\cite{tinkham:1963,tinkham:1964}
with $\Hcpl=5.6$\,T and $\Hcpd=1.1$\,T.}
\label{tinkham}
\end{figure}

An estimate of the thickness of the superconducting sheet can be retrieved by 
measuring the magneto-transport properties of our BLs. Such measurements are 
evaluated in the Ginzburg-Landau (GL) regime for magnetic 
fields up to \Hcc. They also act as an 
independent proof for the observed 2D superconductivity given that the sheet 
thickness \dsc is comparable to or smaller than the GL coherence length 
$\xi_\mathrm{GL}$ \cite{tinkham:1996,degennes:1999}.
Fig.~\ref{tinkham} exemplarily summarizes the observed dependence of the
upper critical magnetic field \Hcc of a BL with $N=60$ on the angle $\theta$, 
in respect to the sample plane of the externally applied swept magnetic field. 
As expected for a 2D superconductor the in-plane field \Hcpl 
($\theta=0$\textdegree) is much higher than the out-of-plane field \Hcpd 
($\theta=90$\textdegree).
Following Tinkham's analysis \cite{tinkham:1963,tinkham:1964,harper:1968} 
both magnetic fields can be experimentally determined on the basis of
the relation:
\begin{equation}
\left|\frac{\Hcc(\theta)\sin\theta}{\Hcpd}\right|+
\left(\frac{\Hcc(\theta)\cos\theta}{\Hcpl}\right)^2=1.
\label{tinkhamrelation}
\end{equation}

The GL theory for thin superconducting sheets allows to express both magnetic 
fields in terms of the magnetic flux quantum $\phi_0$, the thickness \dsc of 
the 2D superconducing sheet, the product of the penetration depth and the 
thermodynamic critical field~\cite{ginzburg:1950}. 
Eliminating last said quantity the thickness \dsc can be 
calculated~\cite{harper:1968} with the help of experimentally obtained values 
of \Hcpl and \Hcpd via
\begin{equation}
\dsc=\sqrt{\frac{6\phi_0\Hcpd}{\pi\Hcpl^2}},
\end{equation}
supporting a thickness of the 2D superconducting sheet to be in the range of 
$\dsc\approx11.8$\,nm being clearly smaller than the nominal thickness of the 
BPO top layer. The calculated thickness \dsc is robust against different 
experimental determination of \Hcc (see Supplemental 
Material~\cite{supplement}, Fig.~S3). 
In contrast, \dsc of a nominal \BBPOF thin film (60~LP) is of the same size as  
the overall thickness of the thin film (see Supplemental 
Material~\cite{supplement}, Fig.~S4).  

The coherence length of the BL ($N=60$) is estimated to be 
$\xi_\mathrm{GL}\approx11.0$\,nm, the value of which we extracted from the 
temperature dependence of the upper critical magnetic field obtained from 
$R(T, H)$ measurements (see Supplemental Material~\cite{supplement}, Fig.~S5). 
This is consistent with the prerequisites of the GL theory~\cite{tinkham:1996}.

\begin{figure}
\centering
\includegraphics[width=\columnwidth]{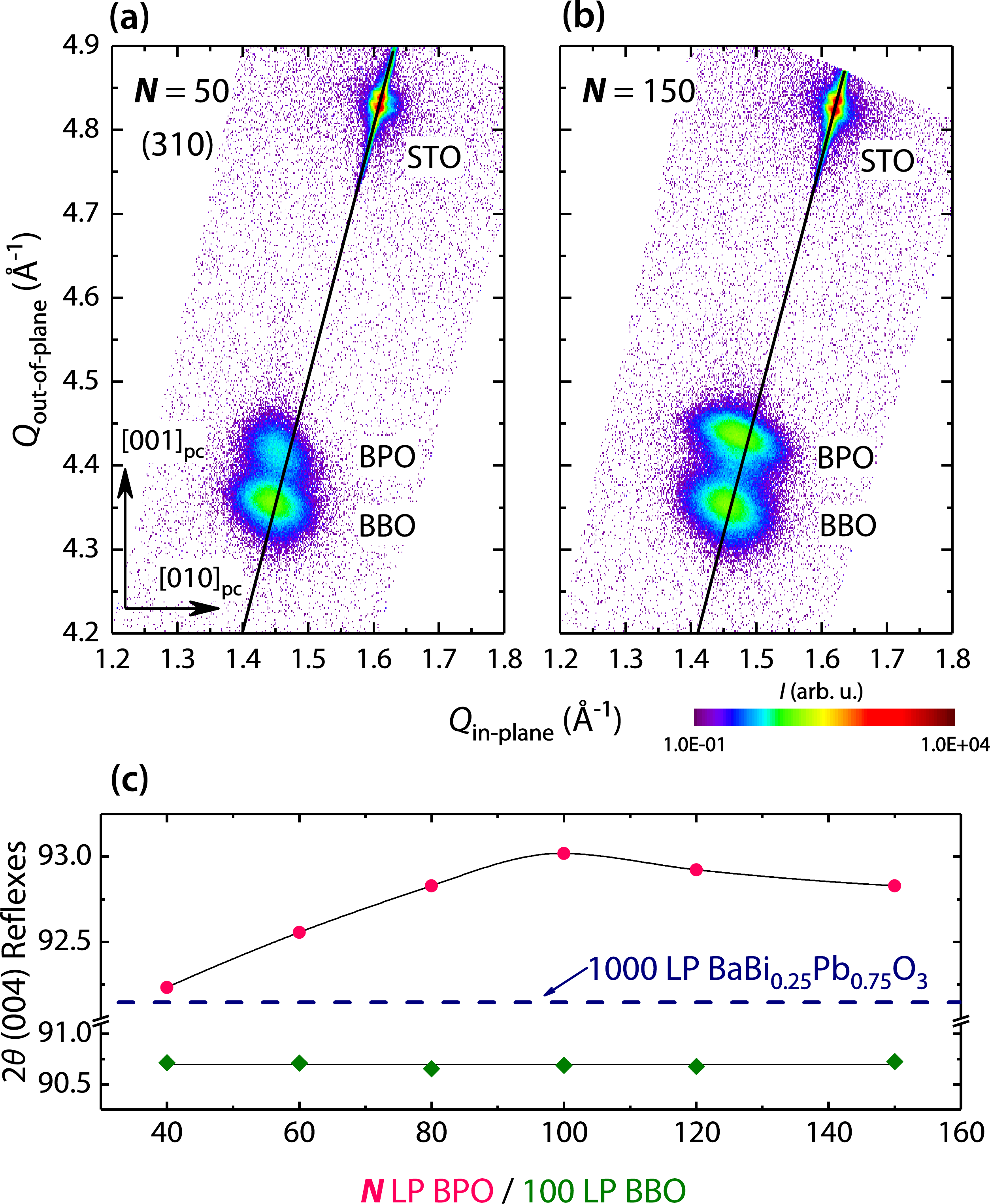}
\caption{XRD studies on BLs of BBO (100~LP)/BPO ($N$~LP) grown on (001) 
oriented STO substrates. Reciprocal space maps of BLs for a) $N=50$ 
(superconducting) and b) $N=150$ (metallic). Whereas BBO grows completely 
relaxed on STO substrates following the relaxation line (black), the locations 
of both peaks of BPO are clear indicators for a strained growth of BPO on BBO. 
The observed shift of the BPO peak away from BBO peak for larger $N$ supports 
the idea that superconductivity is strain-induced. c) Obtained (004) peak 
positions of BBO and BPO in BBO/BPO BLs from a series of $\theta/2\theta$ scans 
(black lines serve as guides to the eye). Whereas BBO shows no significant 
shift in its peak positions, the BPO peaks move to lower angles with 
decreasing $N$ supporting on-going strain-induced growth.}
\label{XRD}
\end{figure}

We suggest that the occurrence of 2D superconductivity in BBO/BPO BLs 
results from a strained growth of BPO on top of BBO possibly also affecting the 
crystallographic phase of BBO at the interface. Reciprocal space maps, which 
were taken for BLs with $N=50$ and $N=150$ (see~Fig.~\ref{XRD}a and 
\ref{XRD}b), prove that BBO grows epitaxially and fully relaxed on STO 
substrates consistent with domain-matched 
expitaxy~\cite{zheleva:1994,narayan:2003}. BPO however grows strained on top of 
BBO for thin layers ($N=50$, Fig~\ref{XRD}a) and relaxes more and more for 
thicker layers ($N=150$, Fig.~\ref{XRD}b). We studied the strain effect in more 
detail in XRD $\theta/2\theta$ measurements taken for several BLs with 
different $N$ (see~Fig.~\ref{XRD}c). With increasing $N$ the $2\theta$ position 
of the (004) peak of BPO shifts towards larger angles supporting a relaxation 
of interface-induced strain in the on-growing lattice of BPO. For BLs with 
$N=\{80, 100, 150\}$ the peak position of BPO stabilizes serving as an 
indication for reaching a relaxed state. 

These observations aid the idea of a strain-only effect at the interface: as 
BLs with $N=50$ become superconducting, whereas those with $N=150$ stay 
metallic in the accessible temperature range, a possible existence of an 
established homogeneously doped region equal to \BBPO is hardly conceivable. 
Ultimately a most favorable thickness of the BPO film exists to assign a 
robustly induced superconducting phase with maximum \Tc.

In this study we have shown that BLs of BBO and BPO become superconducting 
depending on the thickness of the BPO top layer with transition temperatures as 
high as $\Tcip=3.5$\,K for our given sample layout. The electrical transport 
characterization clearly revealed a BKT transition identifying a 2D 
superconducting state within the BLs. Magneto-transport investigations 
independently provide evidence for the two-dimensionality of the 
superconducting sheet estimating the thickness of the sheet to be 
$\dsc\approx11.8$\,nm. X-ray studies suggest an interface-driven 
epitaxial-matched strain effect at the interface between BBO and BPO 
responsible for the occurrence of the observed 2D superconducting state.
The encountered transition in BBO/BPO bilayers is a remarkable further 
precedent of interface driven ground-state variance. With the identification
of a robust superconducting state at the interface of two ``hidden'' topological
insulators~\cite{li:2015} it will be compelling to determine the surface states 
of the superconducting sheet in future work.

\begin{acknowledgments}
We thank A.\,P.~Kampf for helpful discussions. We thank S.~Esser for helpful 
discussions and his support in the measurement and the evaluation of the data 
for the reciprocal space maps. This work was supported by the DFG through TRR\,80.
\end{acknowledgments}

\bibliographystyle{apsrev}
\bibliography{hammerl}
\end{document}


\pagestyle{scrheadings}
\chead{\normalfont Supplemental Material}
\cfoot{\footnotesize B. Meir, S. Gorol, T. Kopp, G. Hammerl,\\[-3mm] Interface-Driven Two-Dimensional Superconductivity in Bilayers of \BBO and \BPO}
\title{Interface-Driven Two-Dimensional Superconductivity\\
	in Bilayers of \BBO and \BPO}

\author{B.~Meir}
\author{S.~Gorol}
\author{T.~Kopp}
\author{G.~Hammerl}
\affiliation{
	Experimental Physics VI\\ 
	Center for Electronic Correlations and Magnetism, Institute of Physics, University of Augsburg, 86135 Augsburg, Germany}

\section*{Supplemental Material}

\begin{abstract}
With this Supplemental Material we provide additional x-ray characterization of our samples and explicitly discuss
the extraction of critical values such as the BKT transition temperature \TBKT, the upper critical magnetic fields \Hcpd and \Hcpl, the thickness
of the superconducting sheet \dsc, and the GL coherence length $\xi_\mathrm{GL}$ from our measured data.
\end{abstract}

\maketitle

\newpage

\begin{figure}[h]
\centering
\includegraphics[width=1\columnwidth]{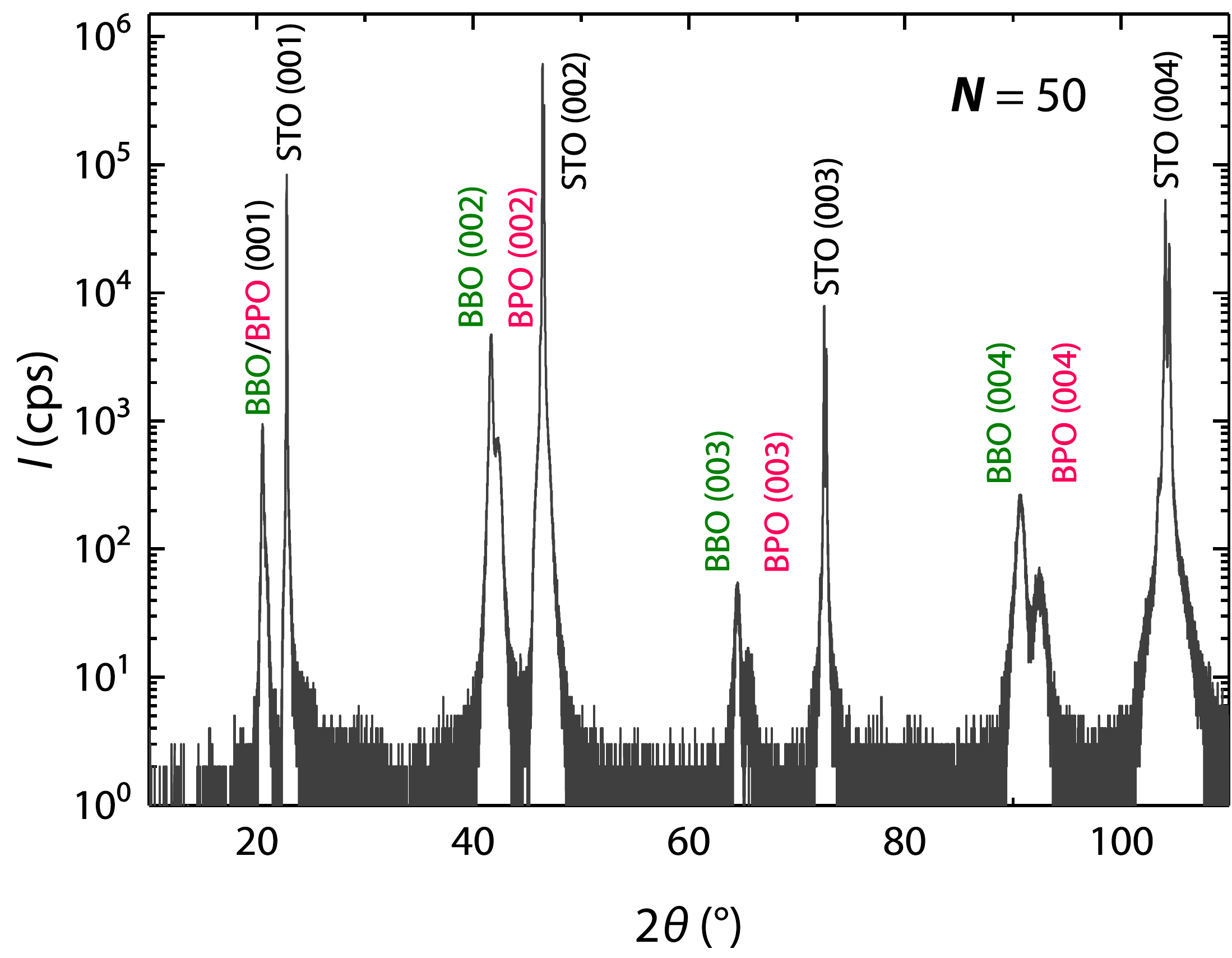}
\caption{Exemplary $\theta/2\theta$ scan obtained from a BBO/BPO BL with $N=50$ (number of laser pulses) grown on (001) oriented STO substrate by PLD. 
The peaks assigned to BBO and BPO are well separated in individual reflexes at their characteristic positions, respectively. Peaks attributed to a homogeneously doped \BBPO phase can not be identified.
}
\label{s1_xrd}
\end{figure}

\newpage

\begin{figure}
\centering
\includegraphics[width=.7\columnwidth]{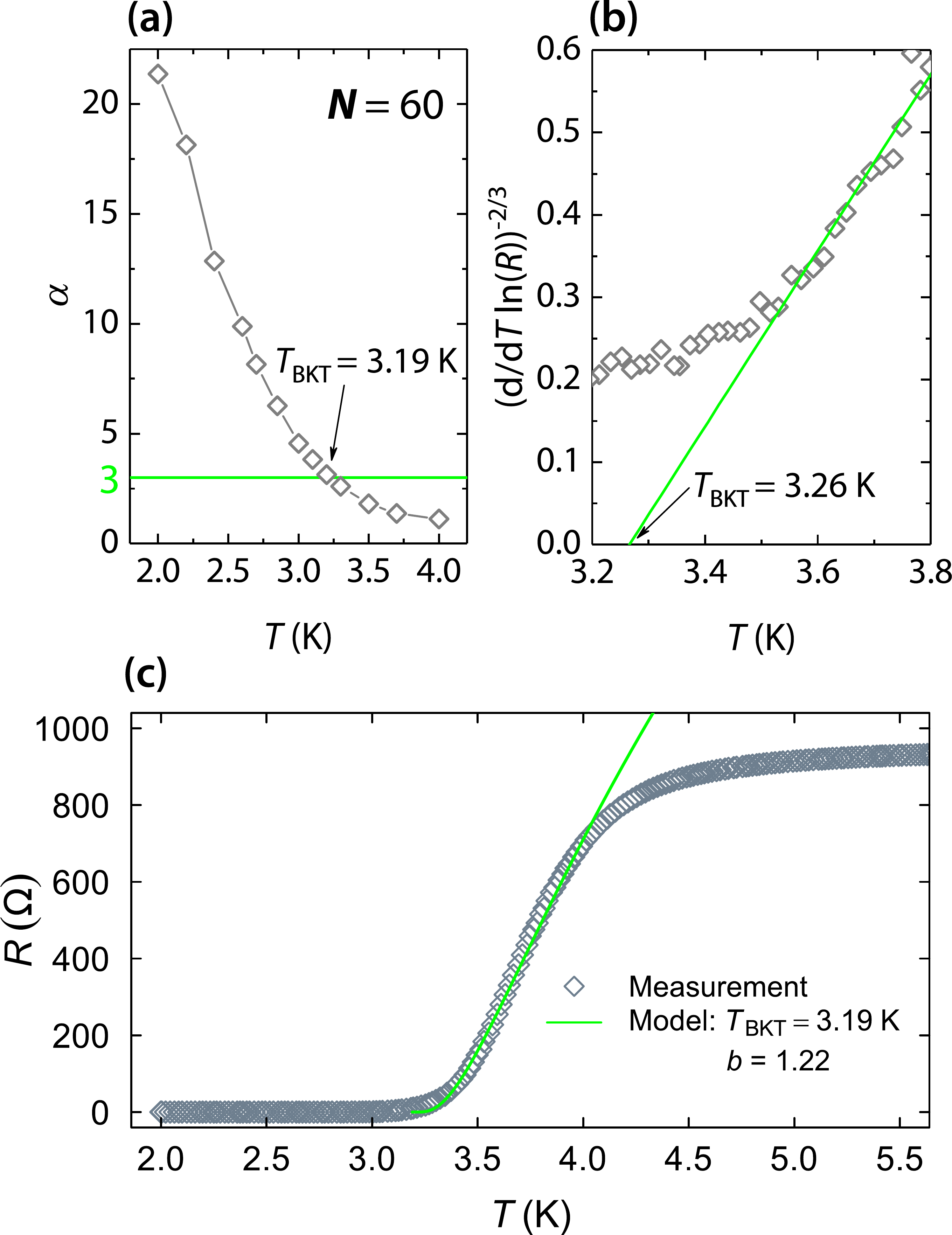}
\caption{Independent evaluation of \TBKT from current-voltage characteristics and the  $R(T)$ dependence of a BBO/BPO BL with $N=60$ (number of laser pulses). 
a) Obtained dependence of the exponent $\alpha$ taken from the IV measurements shown in Fig.~2 as a function of temperature. 
The BKT transition is identified where $\alpha=3$ resulting in $\TBKT=3.19$\,K. b) \TBKT can be retrieved from the
$R(T)$ dependence where the resistance is described by $R\propto R_\mathrm{n}\exp(-b/\sqrt{t})$ with $b$ being a material parameter in the range of unity and $t=T/\TBKT-1$. 
A fit to the data yields $\TBKT=3.26$\,K. c) Direct
evaluation of the $R(T)$ curve in terms of the standard BKT formalism yields $\TBKT=3.19$\,K and $b=1.22$.
}
\label{s2_bkt}
\end{figure}

\newpage

\begin{figure}
\centering
\includegraphics[width=0.5\columnwidth]{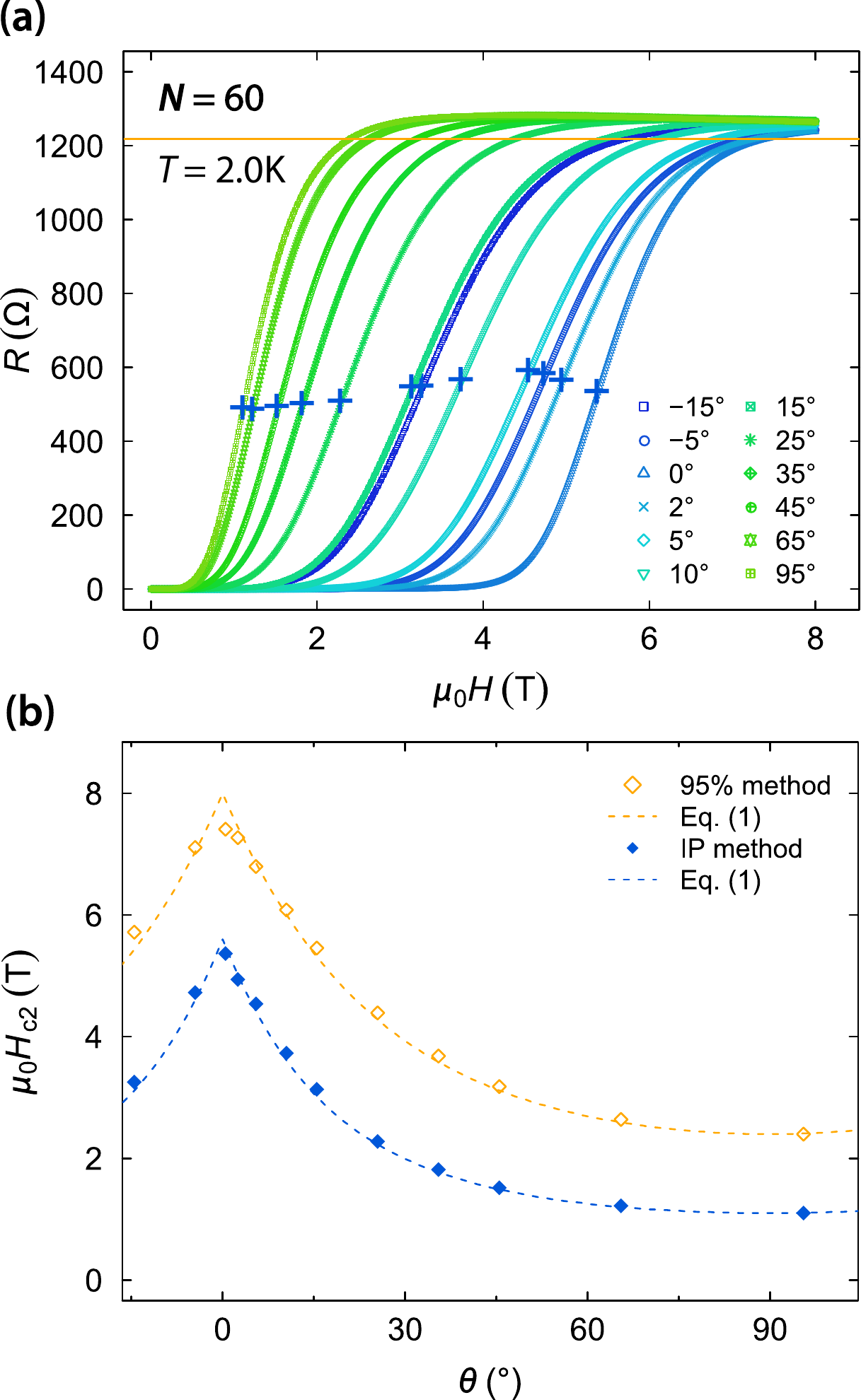}
\caption{Evaluation of \Hcc from magneto-transport measurements taken at $T=2.0$\,K. a)~Dependence of the resistance on the strength of an applied magnetic field in various orientations in respect to the sample plane. The value of
\Hcc is either retrieved by identifying the inflection point (IP) of the $R(H)$ dependence (blue crosses) or by reaching 95\% of the normal-state resistance (indicated by the orange coloured line,
in accordance with the evaluation given, e.\,g., in J.\,M.\,E.~Geers, C.~Attanasio, M\,B.\,S.~Hesselberth, J.~Aarts, and P\,H.~Kes, Phys. Rev. B \textbf{63}, 094511 (2001)).
b)~Obtained dependence of the upper critical magnetic field \Hcc as function of the angle~$\theta$ of the applied swept magnetic field in respect to the sample plane for both evaluation methods described in a). 
The different values of \Hcpl and \Hcpd extracted from both methods (using Equation (1)) result in values of \dsc being 11.8\,nm (blue coloured) and 12.2\,nm (orange coloured), respectively, proving the
robustness of \dsc against different evaluations of the magneto-transport data.}
\label{s3_tinkham}
\end{figure}

\newpage

\begin{figure}
\centering
\includegraphics[width=1\columnwidth]{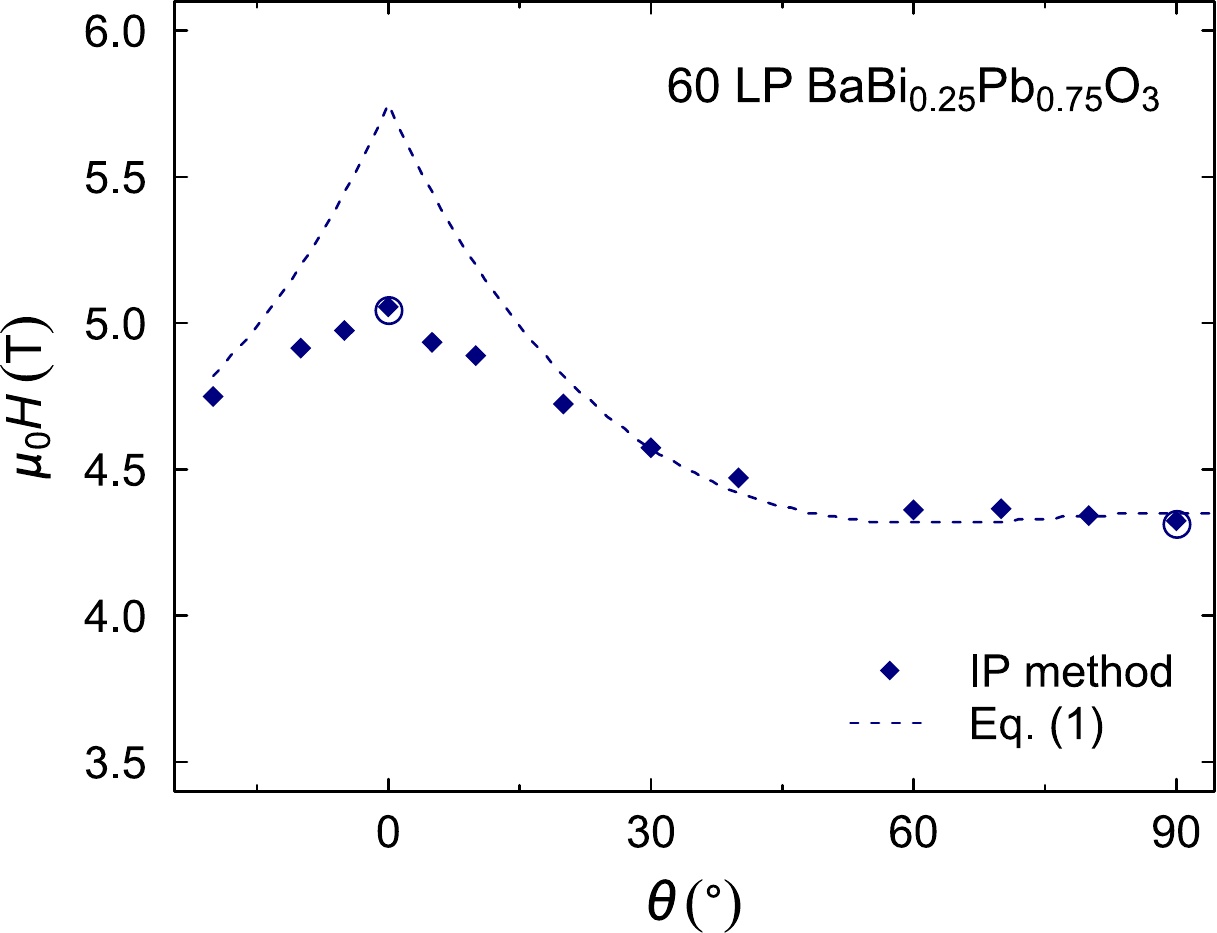}
\caption{Obtained dependence of the upper critical magnetic field 
as a function of the angle~$\theta$ of the applied 
swept magnetic field in respect to the sample plane 
for a homogeneous doped \BBPOF thin film grown by PLD 
with 60~LP and measured at $T=2.0$\,K. The data show a 
reduced influence of the magnetic field on the value of \Hcc
as expected for thicker films (see Ref.~[51]). Extracted values (blue coloured, outlined circles) of
\Hcpl and \Hcpd result in $\dsc=25.9$\,nm comparable to the
film thickness of 24.7\,nm.	
}
\label{s4_bbpo}
\end{figure}

\newpage

\begin{figure}
\centering
\includegraphics[width=0.6\columnwidth]{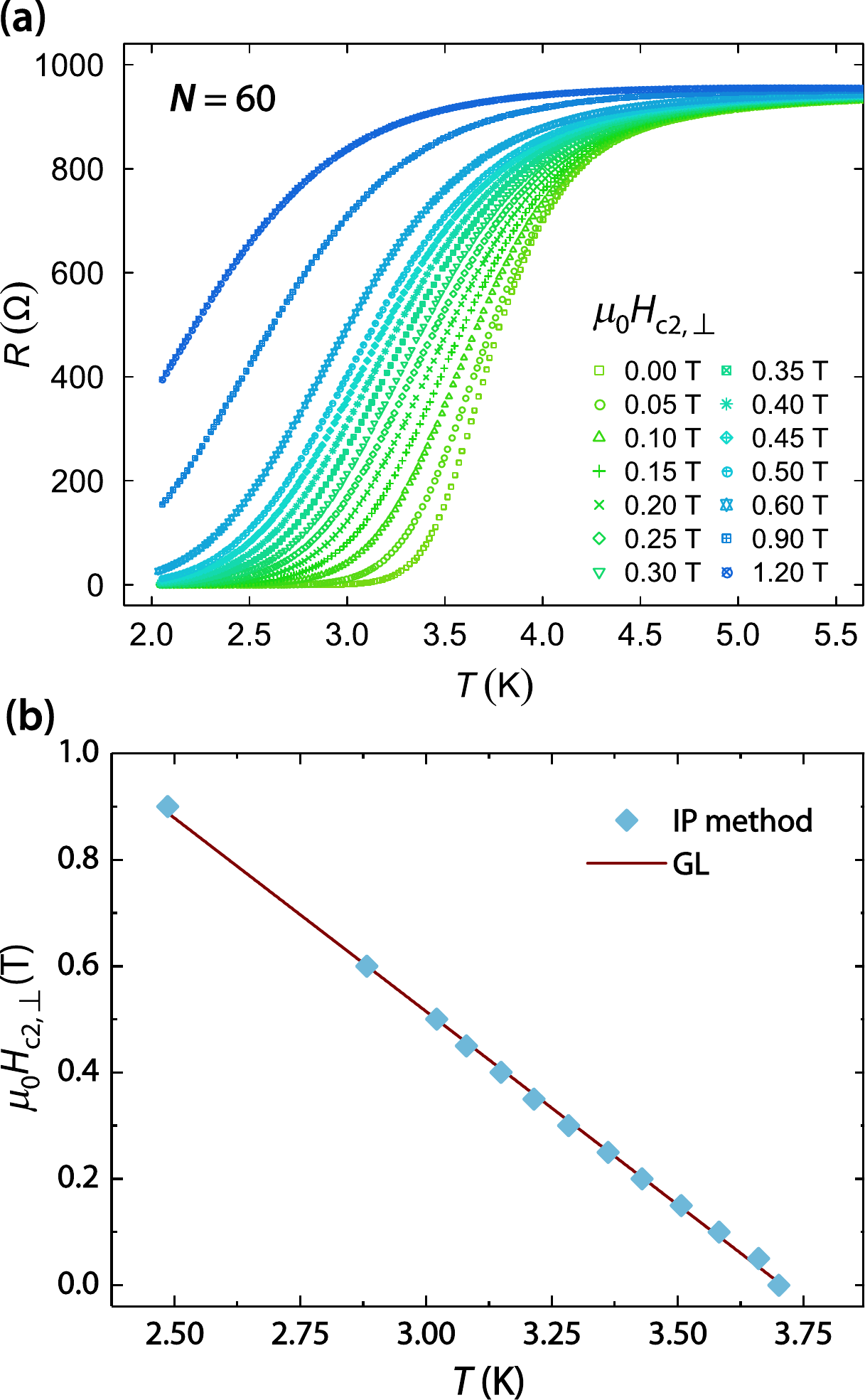}
\caption{Evaluation of the GL coherence length in a BBO/BPO BL with $N=60$. 
a) Dependence of the resistance on the temperature at various magnetic fields applied perpendicular to the sample plane. The value of
\Hcpd is identified at the inflection point (IP) of the magneto-transport curves.
b) Dependence of the upper critical magnetic field \Hcpd as a function of temperature. 
Using $\Hcpd(T)=\frac{\phi_0}{2\pi\xi_\mathrm{GL}^2}(1-T/\Tc)$ the coherence length is calculated to $\xi_\mathrm{GL}=11.0$\,nm.
Using the same criteria for the evaluation of \Hcpd of 95\% of the normal-state resistance as explained in detail in Fig.~S3 the coherence length is
calculated to 9.6\,nm.
}
\label{s5_GL_perp}
\end{figure}